\documentclass[reprint,amsmath,amssymb,aps,prb]{revtex4-1}

\usepackage{graphicx,bm,amssymb,amsmath,dcolumn,hyperref}
\pdfoutput=1
\usepackage{color,multirow}
\usepackage{mathrsfs}
\DeclareMathOperator{\Tr}{Tr}

\begin{document}
\definecolor{red}{rgb}{1,0,0}
\newcommand{\red}[1]{\textcolor{red}{#1}}

\preprint{APS/123-QED}

\title{Fermionic Sign Structure of High-order Feynman diagrams in a Many-fermion System}

\author{Bao-Zong Wang$^{1}$} 
\author{Peng-Cheng Hou$^{1}$}%

\author{Youjin Deng$^{1}$}

\author{Kristjan Haule$^{2}$}

\author{Kun Chen$^{2,3}$}
\email{kunchen@flatironinstitute.org}

\affiliation{$^{1}$ National Laboratory for Physical Sciences at Microscale and Department of Modern Physics, University of Science and Technology of China, Hefei, Anhui 230026, China}
\affiliation{$^{2}$ Department of Physics and Astronomy, Rutgers,
The State University of New Jersey, Piscataway, NJ 08854-8019 USA}

\affiliation{$^{3}$ Center for Computational Quantum Physics, Flatiron Institute, 162 5th Avenue, New York, NY 10010. The Flatiron Institute is a division of the Simons Foundation}

\date{\today}

\begin{abstract}

The sign cancellation between scattering amplitudes makes fermions different from bosons. We systematically investigate Feynman diagrams' fermionic sign structure in a representative many-fermion system---a uniform Fermi gas with Yukawa interaction. We analyze the role of the crossing symmetry and the global gauge symmetry in the fermionic sign cancellation. The symmetry arguments are then used to identify the sign-canceled groups of diagrams. Sign-structure analysis has two applications. Numerically, it leads to a cluster diagrammatic Monte Carlo algorithm for fast diagram evaluations. The new algorithm is about $10^5$ times faster than the conventional approaches in the sixth order. Furthermore, our analysis provides important hints in constructing the relevant effective field theory for many-fermion systems.
\end{abstract}

\maketitle


\section{\label{sec:level1}Introduction}

In quantum mechanics, the amplitudes of fermion-fermion scattering satisfy an anti-commutation relation: permuting two incoming (or outgoing) fermions flips the amplitude's global sign. On the one hand, the sign cancellation plays a vital role in the collective behavior of fermions (For example, it leads to the electron degeneracy pressure which prevents matter from collapsing into a single nucleus\cite{dyson1967stability, lenard1968stability}). 
On the other hand, the same sign cancellation leads to the notorious sign problem in Monte Carlo simulations of many-fermion systems~\cite{PhysRevB.41.9301,PhysRevLett.94.170201}. Indeed, a Monte Carlo estimator of physical observable is a sum of various elementary scattering amplitudes. Even if each amplitude is precisely calculated, the sign-canceled signal can be easily overwhelmed by numerical noise. This paper explores the sign structure of scattering amplitudes in Feynman diagrams to better understand the fermionic sign cancellation.

\begin{figure}
\centering
    \includegraphics[scale=0.55]{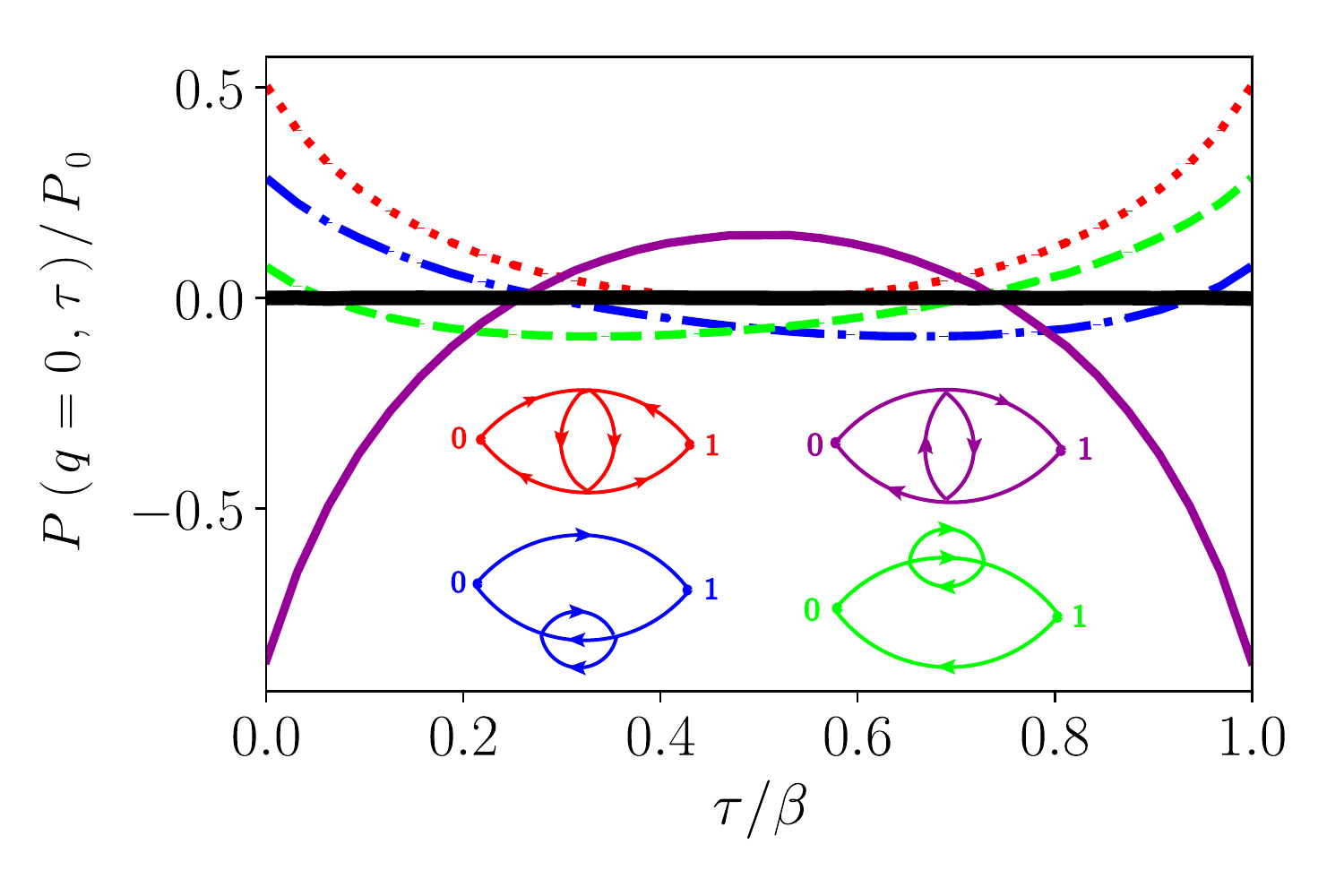}
    \vspace*{-5mm}
    \caption{The imaginary-time evaluations of four third-order Hugenholtz diagrams, which contributes to the static and uniform polarization of a two-dimensional homogeneous ``electron" gas against an external electric-field perturbation. Summing all four diagrams restores the charge conservation law which is broken in each diagrams. At the temperature $T=0.2T_F$, The polarization amplitude drops three orders of magnitude (from $\sim 1$ to $3.48(1)\times 10^{-3}$ 
    in the unit of the free-electron polarization $P_0(q=0, \omega=0)$) as a result of massive sign cancellation between the diagrams.}
     \label{4_diagram_group}
\end{figure}

In many-body physics, Feynman diagrams describe how excitations in the system scatter incoming particles. The diagram amplitude, namely the scattering amplitude, is obtained by summing all diagram topologies and integrating all internal space-time variables (where the excitations get created and annihilated). Both steps involve massive sign cancellations so that the physical scattering amplitude is merely a small fraction of the total diagram weights. Here we use an example to illustrate the non-triviality of this problem. We consider a two-dimensional homogeneous ``electron gas" with Yukawa interaction (see Sec.\ref{sec:model} for the definition of the Hamiltonian). We calculate four diagrams (see Fig. \ref{4_diagram_group}), which are the third-order contributions to the polarization $P$ (namely, the system's linear response function with respect to a screened external electric field perturbation). They all involve a virtual particle-hole pair but with different topologies. As shown in Fig.\ref{4_diagram_group},  their amplitudes are functions of the external imaginary-time. Four diagrams strongly cancel each other: the summed polarization (the thick black line) is merely $0.1\%$ of individual diagrams. Later we show such cancellation is a direct consequence of the emergent charge conservation law in this diagram group---while each diagram is time-dependent, their sum turns out to be a constant of motion, reflecting the conservation of total charge in this system.

The above example leads to fundamental questions we want to address: How does the cancellation occur in diagrams? Does it originate from the summation of topologies or the integration of internal variables? Can one identify the diagram groups that feature massive sign cancellations? Concrete answers to the above questions are important at least in two aspects: i) In fermionic systems, individual-diagram-based power-counting estimation could be misleading in many cases due to the cancellations in diagrams. The sign-canceled-group-based estimation fixes this problem (See. Ref. \cite{belitz1997nonanalytic} and Fig.\ref{4_diagram_group} for examples). ii) Numerically, the knowledge of sign-canceled diagram groups alleviates the sign problem in Monte Carlo techniques, particularly the diagrammatic Monte Carlo methods~\cite{PhysRevLett.81.2514}. In this approach, one calculates the physical observable in terms of diagrammatic expansions with Monte Carlo samplings. The methods have been applied to solve a series of important problems in the Hubbard model~\cite{Kozik_2010,PhysRevLett.113.195301,deng2015emergent,PhysRevB.96.041105,PhysRevB.96.081117,PhysRevB.97.085117,PhysRevB.99.035120,PhysRevB.100.121102,PhysRevB.101.075113,PhysRevLett.124.117602}, many-electron problem~\cite{van2012feynman,PhysRevB.99.035140,PhysRevLett.121.130405,PhysRevLett.121.130406}, Fermi polaron problem~\cite{PhysRevLett.81.2514,PhysRevB.62.6317,Prokof2008Fermi,PhysRevB.77.125101,PhysRevB.90.104510,PhysRevLett.113.166402,PhysRevB.97.134305,PhysRevLett.123.076601,PhysRevB.101.045134} and frustrated spin systems~\cite{PhysRevLett.110.070601,PhysRevB.87.024407,PhysRevLett.116.177203}. The conventional approaches stochastically sample individual diagrams and therefore suffer from the severe sign problem caused by the massive sign cancellation between the diagrams. Recently, a new generation of the algorithms, which sample the summed weight of groups of diagrams, has been developed ~\cite{rossi2017determinant,PhysRevLett.121.130405,KunChen2019,PhysRevB.101.125109,PhysRevB.101.075113}. We will refer them to cluster DiagMC algorithms. With the proper grouping of the diagrams, those algorithms can dramatically reduce the sign problem. Indeed, a couple of cluster DiagMC algorithms are reported to be applicable on much higher-order diagrams~\cite{rossi2017determinant}, or in problems beyond the ability of the conventional approaches\cite{KunChen2019,PhysRevB.101.125109,taheridehkordi2020algorithmic}. Despite the impressive progress, we still lack the design principle on diagrammatic groups to minimize the sign problem.

In this paper, we investigate two universal mechanisms of the sign cancellation, one of which is imposed by the crossing symmetry of the two-body scattering. The other is imposed by the global $U(1)$ symmetry or the charge conservation law, as previously discussed. We show that the symmetries constrain the sign-canceled diagram groups' structure and fix the internal variables' arrangements. We numerically test the idea by calculating the diagrams up to the sixth order for a representative many-fermion system---the uniform Fermi gas with Yukawa interaction. We observe massive sign cancellations in the high-order diagrams and find a significant portion of the overall cancellation originates from the summation of topologies. We then utilize the grouping scheme to formulate a highly efficient cluster DiagMC algorithm. Compared with the conventional approaches, our new algorithm improves the efficiency by $\sim 10^5$ at the sixth order.

We have made three contributions in this paper: i) We prove that in the imaginary-time and momentum representation, the charge conservation law is a topological property of Feynman diagrams. We then clarify its essential role in the diagrammatic sign structure.  ii) We develop the protocol to quantify the sign cancellations from different mechanisms in high-order diagrams. It enables a systematic investigation of the sign-cancellation effects caused by the two symmetries. iii) A more efficient cluster DiagMC algorithm is proposed. Thanks to the symmetry analysis, we can optimize the grouping scheme systematically. The sign problem considerably is significantly alleviated as compared to the algorithm proposed in Ref.\citenum{KunChen2019}.

We now sketch the content of the paper and outline how it is organized. In Sec. II, we describe a two-dimensional uniform Fermi gas with Yukawa potential, which is used to quantitatively investigate Feynman diagrams' sign structure. In Sec. III, we mathematically formulate the concepts of overall and topological sign cancellation. We also clarify the relation between the topological sign cancellation and the efficiency of a typical cluster DiagMC algorithm. Sec. IV and V are dedicated to crossing symmetry and global $U(1)$ symmetry induced sign cancellations. The topological sign cancellation and efficiency of the cluster DiagMC are quantitatively investigated. Finally, we outline our conclusions and discuss the outlook for future development and applications in Sec. VI.

\section{Interacting fermions model}
\label{sec:model}
\begin{figure}
    \includegraphics[scale=0.38]{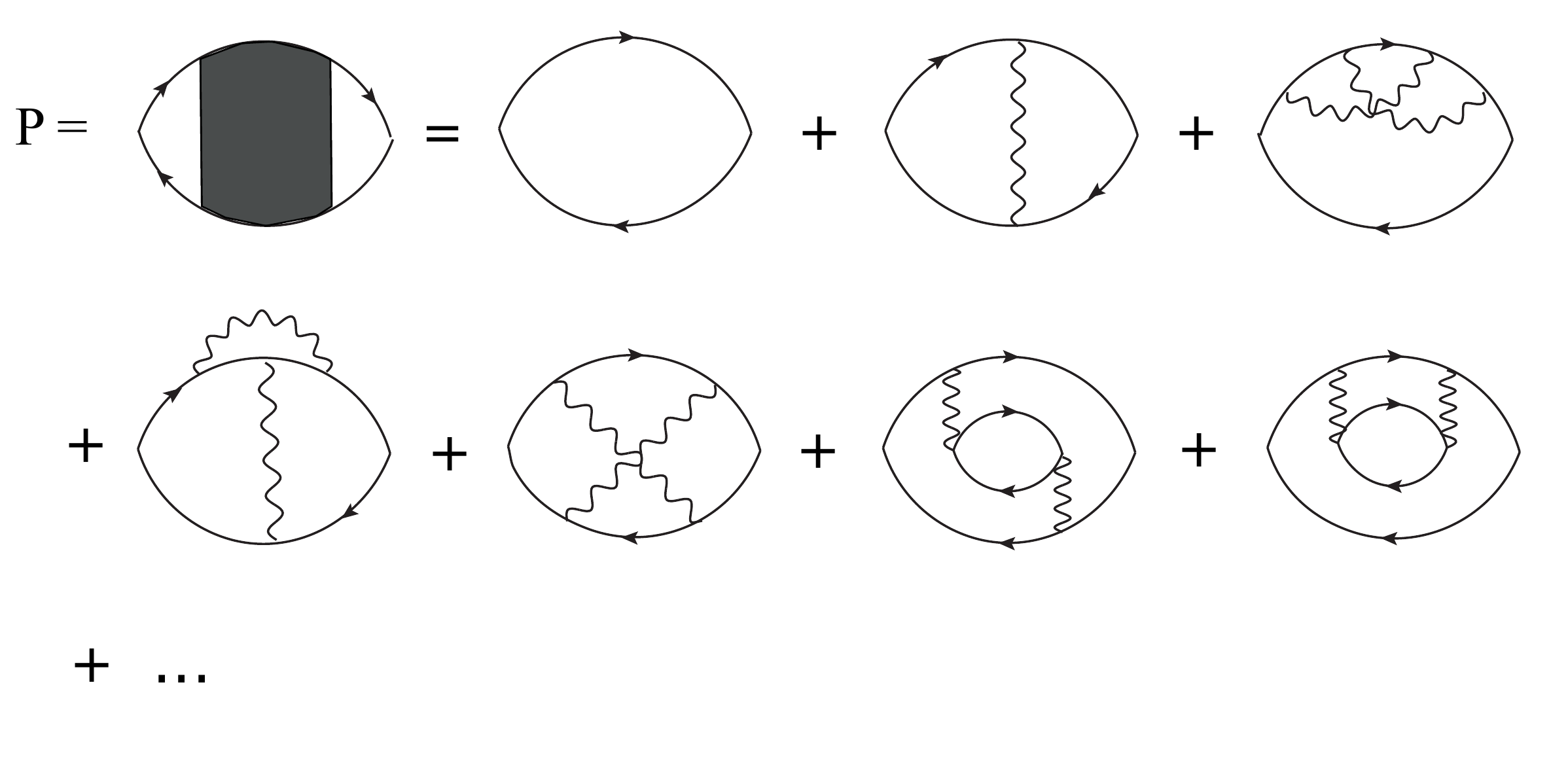}
    \caption{Some examples of the polarization Feynman diagrams. The interaction line represents the bare Yukawa interaction. The Green's function already resumes the Hatree and Fock self-energy. Therefore, all diagrams contains the Hatree and Fock sub-diagrams should be excluded.}
    \label{Fig:polarization}
\end{figure}
To study the sign structure of Feynman diagrams at a quantitative level, we consider a representative model, which is a 2D uniform spin-polarized interacting Fermi gas described by the Hamiltonian,
\begin{equation}
    \hat{H} = \int\frac{d\bm{k}}{(2\pi)^2} \epsilon_k \hat{c}^{\dagger}_{\bm{k} } \hat{c}_{\bm{k} } 
   + \int\frac{d\bm q d\bm{k} d\bm{k'}}{(2\pi)^6} V(q)
    \hat{c}^{\dagger}_{\bm{k}+\bm{q} } \hat{c}^{\dagger}_{\bm{k'}-\bm{q} } \hat{c}_{\bm{k'}} \hat{c}_{\bm{k}},
\label{toymodel}
\end{equation}
where $\hat{c}_{\bm k}^\dagger (\hat{c}_{\bm k})$ is the creation (annihilation) operator of a fermion with momentum $\bm k$, $\epsilon_k=k^2/(2m)-\mu$ is the kinetic energy shifted with the chemical potential $\mu$. $V(q)=4 \pi e^2/(q^2 + \lambda^2)$ is the Yukawa interaction between fermions with $e$ the electric charge and $\lambda^{-1}$ the static screening length. The fermions in this model can be understood as the electrons with a renormalized Coulomb interaction\cite{KunChen2019}. We adapt the natural units $\hbar=c=\varepsilon_0=k_B=1$. With a given screening length $\lambda^{-1}$ and temperature $T$, the physics of the system is controlled by the characteristic length scale, namely the Wigner–Seitz radius $r_s=(1/\pi n)^{1/2}$ with $n$ the density of the fermions.

Hereby we focus on the physical observable polarization $P$, which is the linear response function of the system with respect to a screened external electric field. It is closely related to the dielectric function, Pauli spin susceptibility and charge compressibility of the system \cite{giuliani2005quantum}. More specifically, one couples the system to a small external electric field $\delta E$,
\begin{equation}
\hat{H}[\delta E] = \hat{H} + \int \frac{d \bm q}{(2\pi)^2} \: \hat{n}_{\bm q} \delta E_{\bm q} ,
\end{equation}
where $\hat{n}$ is the charge density operator. We assume the external electric field doesn't change with the time since we are mainly interested in the static polarization. The generating functional for the connected correlation function is the grand potential of the system $\Phi[\delta E]=\ln \Tr \exp \{-\beta \hat{H}[\delta E]\}$. In particular, the two particle correlation function (charge susceptibility) $\chi$ is given by,
\begin{equation}
\chi(\bm x_1-\bm x_2; \omega=0) = \frac{\delta^2 \Phi[\delta E]}{\delta E({\bm x}_1) \delta E({\bm x}_2)}\bigg|_{\delta E\to 0},
\label{Eq:Gchi}
\end{equation} 
where the indexes ${\bm x}_1$ and ${\bm x}_2$ are the spatial coordinates of the external perturbation. The polarization is the charge response function with respect to the screened external electric field, rather than the original field $\delta E$. It can be derived from the charge susceptibility,
\begin{equation}
P(\bm q, \omega=0) = \frac{\chi(\bm q, \omega=0)}{1+V(\bm q)\chi(\bm q, \omega=0)}.
\label{polarization}
\end{equation}

When the fermionic interaction $V(\bm q)$ is relatively weak, the polarization can be effectively calculated with the Feynman diagrammatic technique. One first performs a double power expansion of the generating functional $\Phi[\delta E]$ in both $V(\bm q)$ and $\delta E$, then derives the power series of the charge susceptibility using Eq.(\ref{Eq:Gchi}). Each term in the power series can be vividly represented by a Feynman diagram. They are connected diagrams composed of bare propagator lines and bare interaction lines. All of them have two external vertices connecting two external perturbations. In the end, the polarization diagrams can be derived from the charge susceptibility diagrams using Eq.(\ref{polarization}). Diagrammatically, it means one should exclude all reducible diagrams that would fall into two pieces if one interaction line were cut off. We show some of the first two orders of polarization diagrams in Fig. \ref{Fig:polarization}. We use the momentum and imaginary-time representation so that all diagram weights are real-valued.

To further improve the convergence of the diagrammatic series, we replace each bare Green's function line in the diagrams with the renormalized Green's function which resumes the Hatree-Fock self-energy,
\begin{equation}
\begin{aligned}
    G(\bm{k},\tau)= e^{-\epsilon_{\bm{k}} \tau} \left[ (1-n_{\bm{k}})\theta(\tau) - n_{\bm{k}} \theta(-\tau) \right],
\label{green}
\end{aligned}
\end{equation}
where $\epsilon_{\bm k}={\bm k}^2/{2m}-\mu-\epsilon_{HF}(\bm k)$ with $\epsilon_{HF}(\bm k)$ the Hatree-Fock self-energy which doesn't depend on the imaginary-time. To avoid double-counting in the diagrammatic series, all diagrams  which contain at least one Hatree-Fock sub-diagram are removed.

\begin{figure}
    \centering
    \includegraphics[scale=0.65]{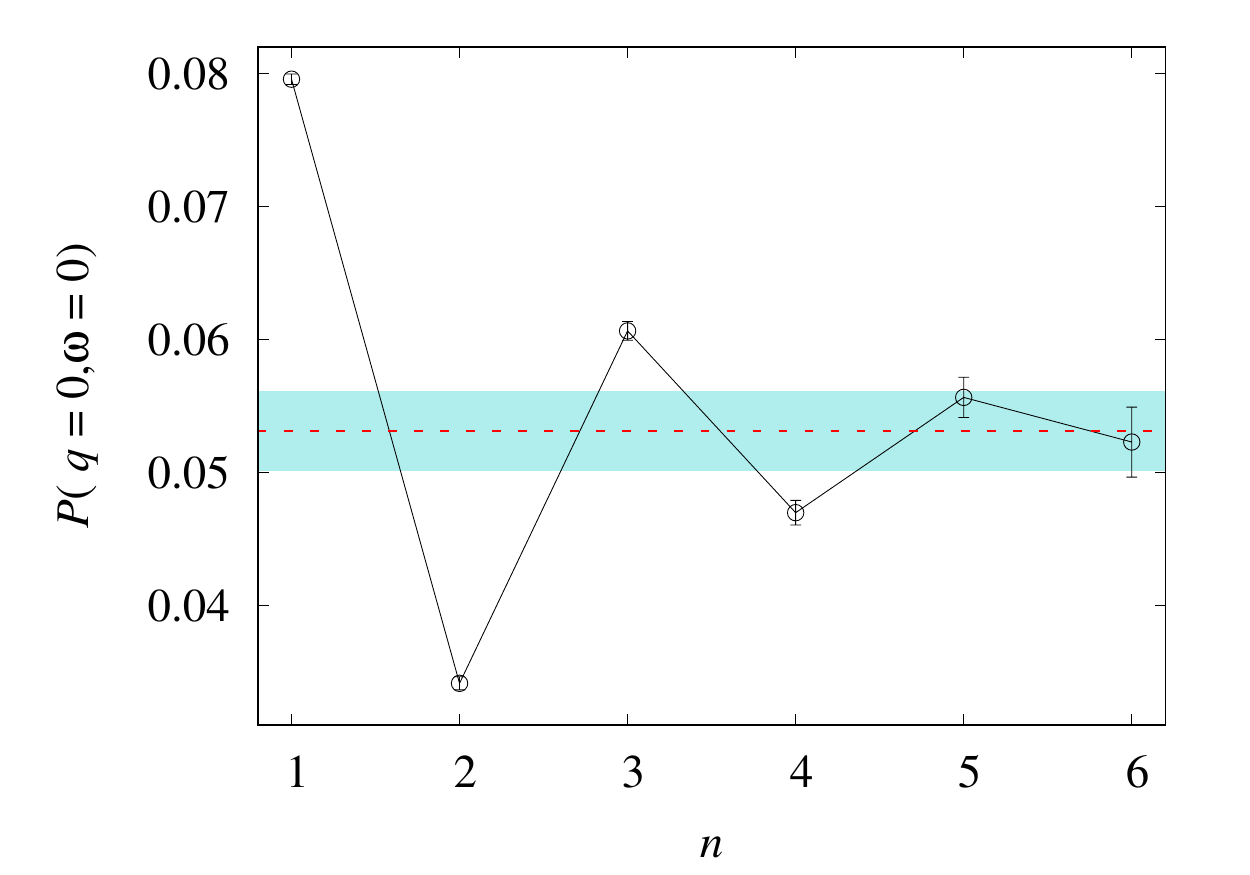}
    \caption{The polarization $P(q=0,\omega=0)$ versus order $n$ for the uniform Fermi gas with Yukawa potential at the temperature $T=0.04T_F$. In the infinite-order limit, the extrapolated polarization is calculated to be $P(q=0,\omega=0)=0.0504(3)$. The expectation value and error bar are marked with the red dashed line and the blue strip, respectively. The latter includes both the statistical error and the extrapolation error.}
    \label{Fig:polarizationQ0}
\end{figure}

In our numerical calculation, we fine-tune the chemical potential $\mu$ so that the Fermi momentum of the Green's function Eq. \ref{green} matches $k_F=(4/S)^{1/2}/r_s$, where the spin factor $S=1$ and the radius $r_s=\sqrt{2}$. As one includes higher-order quantum corrections, we expect the Fermi momentum, as well as the Wigner–Seitz radius, get renormalized. Therefore, both $k_F$ and $r_s$ should be slightly different from the actual physical values. It does not affect the claims in this paper.
 At the temperature $T=0.2E_F$, we calculate static polarization diagrams $P^{(n)}(q=0,\omega=0)$ 
 up to order $n=6$. The result is shown in Fig.~\ref{Fig:polarizationQ0}. With the above parameters, we observe exponential convergence with the truncated diagram order.

\section{Overall and Topological Sign Cancellation}
\label{sec:sign_cancellation}
This section quantifies the concepts of overall and topological sign cancellation in Feynman diagrams. We also clarify the connection between topological sign cancellation and cluster diagrammatic Monte Carlo algorithms' efficiency.

We consider a diagrammatic series for a positively defined physical quantity $F$ (for example, the polarization Eq.\eqref{polarization}),
\begin{equation}
F^n =\int[{\rm d}\bm x]^{n}\sum_{\xi_n}f(n,\xi_n,\bm x), 
\label{F}
\end{equation}
where $f(n,\xi_n,\bm x)$ is the weight of an $n$-th order diagram with a topological index $\xi_n$ and a set of internal variables $\bm x$ (in this paper, they are the internal momenta and imaginary times). Due to the sign cancellation between diagrams, we expect $F^n$ to be much smaller than the weight function,
\begin{equation}
F^n_{\rm abs} =\int[{\rm d}\bm x]^{n}\sum_{\xi_n} \left|f(n,\xi_n,\bm x)\right|,
\label{F_abs}
\end{equation}
where sums the absolute weight of the diagrams. 

\begin{figure}
    \includegraphics[scale=0.6]{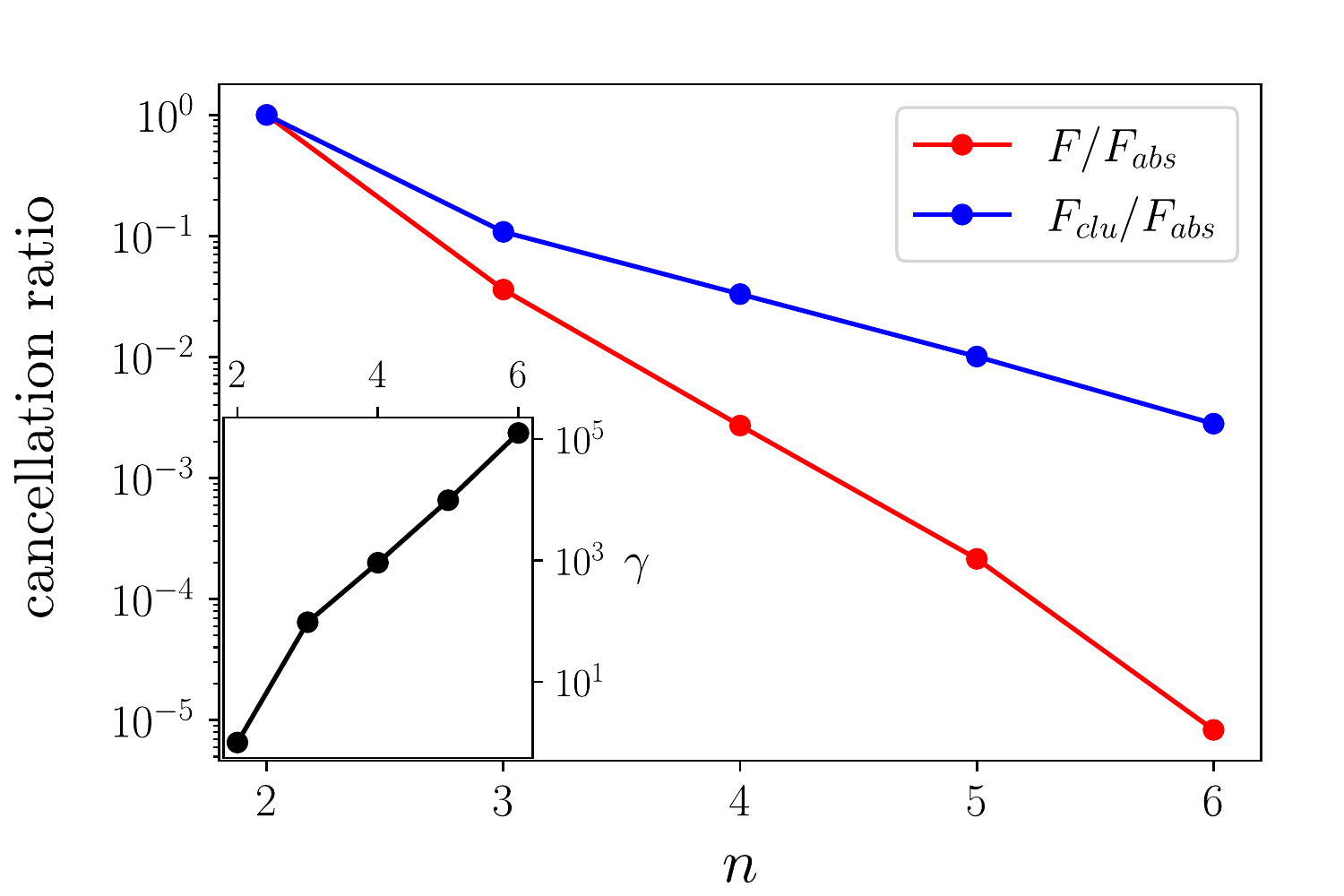}
    \caption{Sign cancellation of the polarization diagrams in a two-dimensional uniform Fermi gas. The ratio $F/F_{\rm abs}$ represents the quantum sign cancellation between the diagrams. The small ratio $F^n/F^n_{\rm abs}\ll 1$ at the high order indicates most of the diagram weights get canceled. The ratio $F_{\rm clu}/F_{\rm abs}$ estimates the sign cancellation caused by the diagram topologies rather than the internal variable integration. Such sign cancellation can be used to improve the sign problem in the cluster diagrammatic Monte Carlo method. The sign-optimized cluster algorithm can be $\gamma=(F_{\rm abs}/F_{\rm clu})^2$ times faster than the conventional algorithm (see the inset).}
    \label{cancellation}
\end{figure}

We numerically calculate the ratio $F^n/F^n_{\rm abs}$, where $F^n$ is the power series of the polarization in Fig.\ref{polarization}. As shown in the red curve in Fig.\ref{cancellation}, we find this ratio roughly exponentially decays to zero with increasing orders, indicating massive sign cancellations. As a result, while the absolute weight function $F^n_{\rm abs}$ factorially diverge (the number of Feynman diagrams diverge as $2^n n!$), the sign-canceled net contribution to the polarization exponentially converges. In literature, this effect is referred as the sign blessing ~\cite{Prokof2008Fermi}. In what follows, we will define the overall sign cancellation as the ratio $F^n/F^n_{\rm abs}$.

We expect a significant amount of the overall sign cancellation can be traced back to the sign cancellation between different diagram topologies before the internal variables are integrated out. To measure the topological sign cancellation, one should classify the diagrams into sign-canceled topological clusters $C_n$, then calculate the absolute weight of the clusters rather than individual diagrams,  
\begin{equation}
\label{F_clu}
F_{\rm clu}^n =\int[{\rm d}\bm x]^{n} \sum_{C_n} \left|\sum_{\mathcal \xi_n \in C_n} f(n,\xi_n,\bm x)\right|.
\end{equation}
This weight function satisfies an exact inequality,
\begin{equation}
\label{inequality}
F^n\le F^n_{\rm clu} \le F^n_{\rm abs}.
\end{equation}
Of course, the magnitude of $F_{\rm clu}^n$ strongly depends on the choice of the topological cluster $C_n$ and the arrangement of internal variables. Our overall goal is to find the optimized topological clusters to make $F_{\rm clu}^n$ as small as possible. We calculate the ratio $F_{\rm clu}^n/F^n_{\rm abs}$ for the polarization using the clustering scheme discussed in the following section. As shown in the blue curve in Fig.\ref{cancellation}, we find the ratio $F_{\rm clu}^n/F^n_{\rm abs}\ll 1$, indicating a significant amount of the overall sign cancellation indeed has a topological origin. 

Now we discuss the relation between the above weight functions and the efficiency of DiagMC algorithms. In conventional diagrammatic Monte Carlo approaches~\cite{PhysRevLett.81.2514,Prokof2008Fermi,VANHOUCKE201095,Kozik_2010,van2012feynman,deng2015emergent}, the weight function Eq.(\ref{F_abs}) is treated as the weight of the configuration space spanned by the toplogies, the internal variables as well as the order of diagrams. It defines a probability density, $\rho(n, \xi_n, \bm{x})=\left|f(n,\xi_n,\bm x)\right|/F^n_{\rm abs}$, which can be efficiently sampled with a Metropolis algorithm. Supplemented with a normalization scheme, one can use the Monte Carlo estimator $f(n,\xi_n,\bm x)/\left|f(n,\xi_n,\bm x)\right|$ to calculate diagram weight $F^n$. 

In contrast to the conventional algorithm, the recent cluster algorithms~\cite{rossi2017determinant,PhysRevLett.121.130405,KunChen2019,PhysRevB.101.125109} propose to use the weight function Eq.\eqref{F_clu} instead of Eq.(\ref{F_abs}). It is reported that the sign problem is greatly alleviated. We derive the following relation to explain such observation,
\begin{equation}
\label{error}
    \epsilon_{\rm MC}(M) \sim Z/\sqrt{M} \;.
\end{equation}
The detailed derivation is presented in Appendix \ref{errorbar}. Here $M$ is the number of Monte Carlo samples, $\epsilon_{\rm MC}$ is the statistical absolute error in a typical set up of DiagMC algorithm, and $Z$ is the weight of the configuration space. To achieve the desired relative accuracy, the number of the samples should be proportional to the squared ratio $\sim (Z/F^n)^2$. Therefore, the smaller $Z$, the less sign problem. Indeed, the conventional DiagMC choose $Z=F^n_{\rm abs}$, while the cluster algorithms choose $Z=F^n_{\rm clus}$. we expect the latter improves the efficiency by a factor,
\begin{equation}
    \gamma=(F^n_{\rm abs}/F^n_{\rm clu})^2.
\end{equation}

To achieve better efficiency, one should minimize the weight function $F^n_{\rm clu}$.
Of course, this is quite challenging in general. In fact, we find that a randomly picked arrangement of internal variables can hardly improve the sign problem, namely $F^n \ll F^n_{\rm clu} \approx F^n_{\rm abs}$. To make $F^n_{\rm clu}$ small, one has to carefully group the topology of diagrams and arrange the internal variables, so that the diagrams with similar topologies maximally cancel with each other before the integration of internal variables $\mathbf{x}$. Thus, we are motivated to explore the underlying mechanism of the topological sign cancellation in the next two sections.

\section{Topological Sign Cancellation}
In this section, we explore how different symmetries lead to the topological sign cancellation between Feynman diagrams. We focus on two most common symmetries: the crossing symmetry in the two-fermion scattering amplitudes, and the $U(1)$ global gauge symmetry associated with the charge conservation.
\subsection{Crossing-Symmetry induced Sign cancellation}
\label{sec:hugenholz}
\begin{figure}
    \centering
    \includegraphics[scale=0.5]{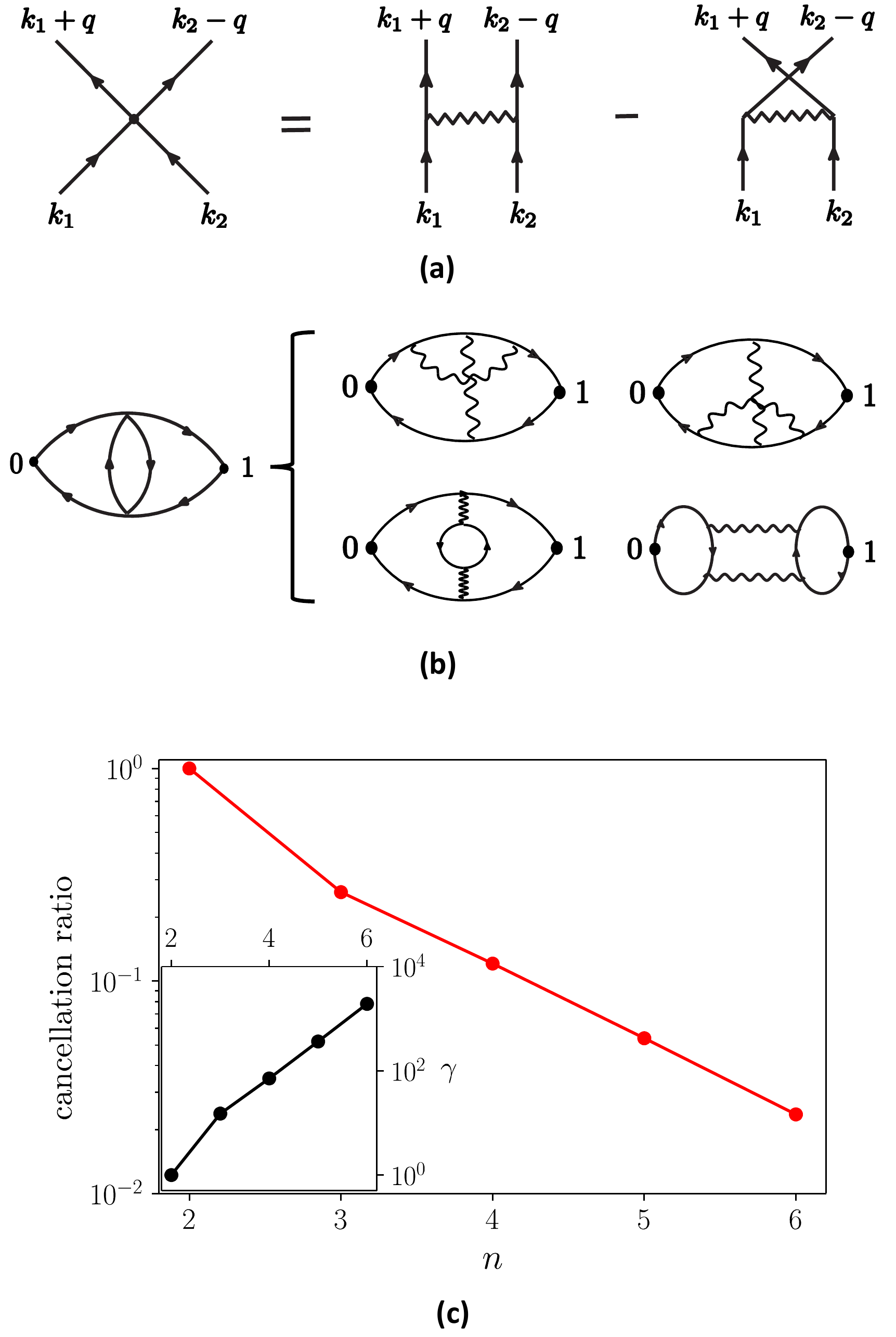}
    \caption{Sign cancellation from the crossing symmetry. (a) Any $4$-point vertex has a direct and an exchange contribution, which always cancels with each other. (b) Combining each pair of direct and exchange interactions into a vertex group $2^n$ Feynman diagrams into a Hugenholtz diagram. (c) The ratio $F_{\rm{Hugen}}/F_{\rm{abs}}$ measures the sign cancellation between the original Feynman diagrams due to the crossing-symmetry. The inset shows the corresponding efficiency improvement factor $\gamma$ of the cluster diagrammatic Monte Carlo algorithm as the diagram order increases.}
    \label{hugenholz}
\end{figure}

We first discuss the crossing symmetry, which means that the direct and exchange parts of the two-fermion scattering amplitude must be anti-symmetric. It indicates massive sign cancellations when the two incoming fermions have similar momenta and frequencies. Fig. \ref{hugenholz}a is the elementary event of the model Eq. (\ref{toymodel}), where two fermions directly interact through the bare Yukawa interaction. For a given Feynman diagram with $n$ interactions, each interaction line has its direct or exchange contributions. The $2^n$ combination defines a topological group of similar or equivalent Feynman diagrams (some may not be polarization diagrams and must be excluded). We expect a massive sign cancellation in the group. In literature~\cite{Hugenholtz_1965}, the combined direct and exchange interactions are labeled as a single vertex, and the above topological group of Feynman diagrams is referred to as a Hugenholz diagram. Fig.\ref{hugenholz}b gives a second-order example.

The weight function which quantifies the crossing-symmetry induced sign cancellation is given by,
\begin{equation}
F_{\rm Huge}^n=\int[{\rm d}\bm x]^{n} \sum_{\mathscr{H}_n} \left| h(n,\mathscr{H}_n,\bm x)\right|,
\end{equation}
where $h(n,\mathscr{H}_n,\bm x)$ is the amplitude of an $n$-order Feynman diagram belonging to the same Hugenholz topology $\mathscr{H}$. The cancellation ratio $F_{\rm Huge}/F_{\rm abs}$ is shown in Fig.\ref{hugenholz}c. We observe an approximately exponential trend of topological sign cancellation as the diagram order increases.

The massive sign cancellation between Feynman diagrams with the same Hugenholz topology can be used to alleviate the sign problem in the conventional DiagMC algorithm\cite{KunChen2019}. As explained in Sec.\ref{sec:algorithm}, one can implement a cluster DiagMC algorithm that samples the Hugenholtz diagrams instead of the Feynman diagrams. In calculations of the polarization for Eq.\eqref{toymodel}, this trick improves the efficiency by a factor $\gamma=(F_{\rm abs}/F_{\rm Huge})^2$, which is depicted in the inset of Fig.\ref{hugenholz}c.

In the end, we would like to discuss the crossing-symmetry-induced sign cancellation between the high-order scattering amplitudes. They are fermion-fermion scattering mediated by the exctations of the system. Diagrammatically, they are $4$-point vertex function with internal structures. Due to the crossing symmetry, permuting two outgoing (or two incoming) legs of the vertex flips the entire diagram's sign. Therefore, it also makes sense to cluster the diagrams associated with this high-order crossing symmetry. Interestingly, we find that the rule is given by the parquet equations~\cite{Diatlov1957,PhysRev.178.1072,Bicker2004}. Physically, the parquet equations are a non-perturbative technique to self-consistently calculate a two-particle vertex function from the irreducible vertex functions. However, there is also a perturbative interpretation: they provide a set of recursive rules to build higher-order vertex diagrams out of lower-order sub-vertex diagrams. The generated diagrams are automatically organized in a hierarchy of $4$-point vertex (sub-)diagrams. Each (sub-)diagram consists of direct or exchange contributions which greatly cancel with each other. It is also possible to generalize the crossing-symmetry clustering for the two-particle scattering to $n$ particles. One then needs to use the Dyson-Schwinger equations~\cite{PhysRev.75.1736,Schwinger452} instead of the parquet equations.

\subsection{$U(1)$-Symmetry-Induced Sign Cancellation}
\label{sec:conservation}

\begin{figure}
    \centering
    \includegraphics[scale=0.38]{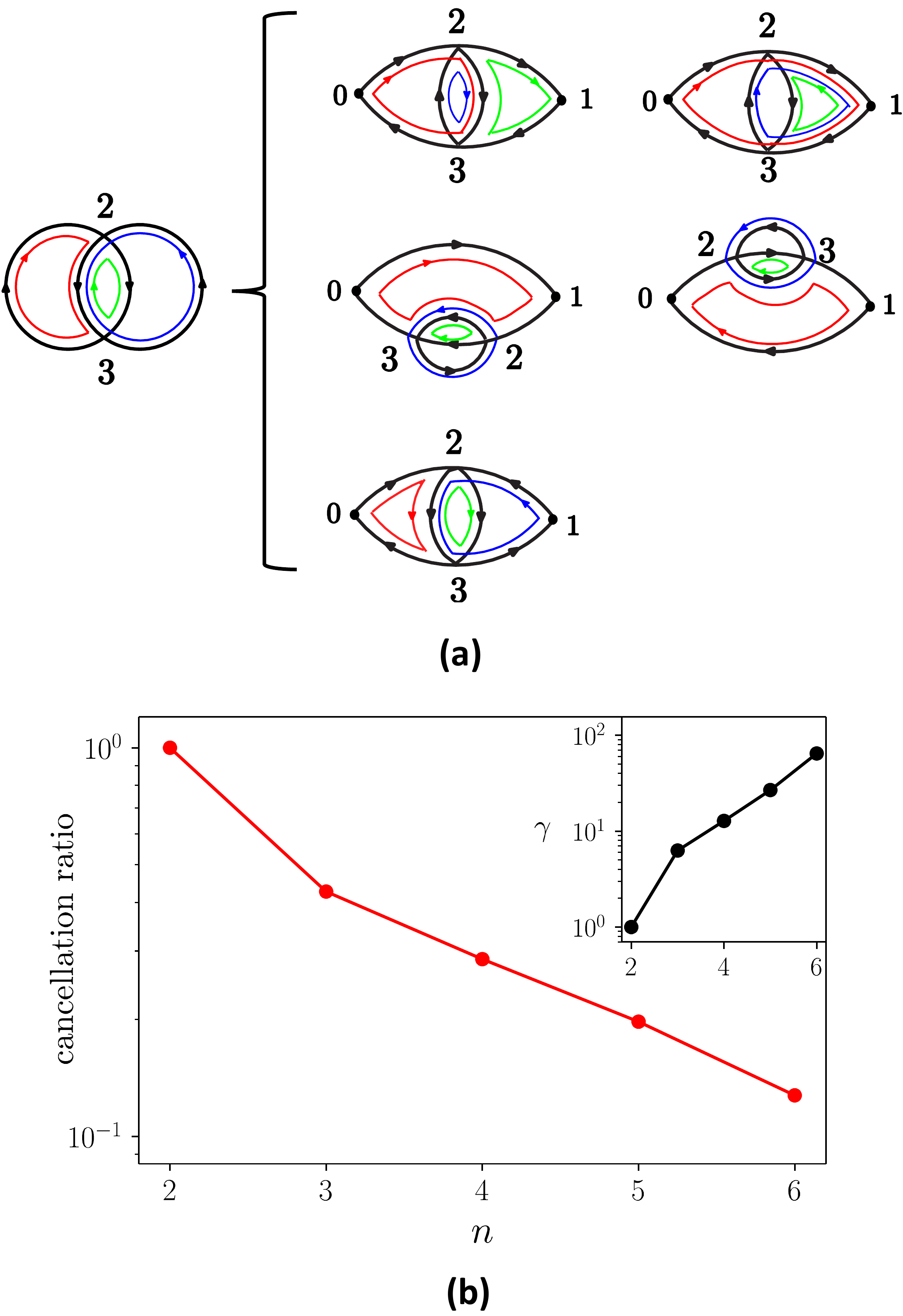}
    \caption{(a) Taking second functional derivatives of one generating functional diagram creates a group of polarization diagrams. Some topologies may be generated multiple times. The copies, as well as the inherited internal momentum loops (the colorful loops with arrows), should be kept as they are. The generated diagram group automatically obeys the charge conservation law before the integral of the internal variables. (b)
    The ratio $F_{\rm{clu}}/F_{\rm{Hugen}}$ measures the sign cancellation between the Hugeholtz diagrams in the conserving group. The inset shows the corresponding efficiency improvement factor $\gamma$ of the cluster diagrammatic Monte Carlo algorithm as the diagram order increases.}
    \label{conservation_law}
\end{figure}

The action of the model (\ref{toymodel}) is invariant under the global transformation $c_{\bm k}\exp(-i\theta)$, where $\theta$ is a space-time independent phase. This global $U(1)$ symmetry, sometimes referred to as the global gauge symmetry, guarantees the conservation of charge (or particle number). It is a property shared by many quantum systems. 

We argue that the total charge conservation leads to a massive overall sign cancellation between diagrams in the introduction. Does such cancellation originate from the summation of diagram topologies? We will approach this problem in two steps: we first show that in the imaginary-time representation, the charge conservation law is a topological property of diagrams, which means that it can be implemented before integrating out the internal variables. We then show that the conserving diagrams inevitably leads to massive topological sign cancellation.

Baym and Kadanoff developed a program to construct conserving approximations using the functional-derivative approach\cite{PhysRev.124.287, PhysRev.124.287}. They showed that any diagrammatic truncation of the generating functional (such as $\Psi[\delta E]$ in Eq.\eqref{Eq:Gchi}) fulfills a set of conservation laws. Taking the derivative of such functional then generates conserving approximations for correlation functions. Their theory explains the emergent charge conservation law in the four polarization diagrams in Fig.\ref{4_diagram_group}: they are all functional derivatives of the same vacuum diagram, as shown in Fig.\ref{conservation_law} (a). The derivative of the vacuum diagram with respect to $\delta E_0$ ($\delta E_1$) splits a Green's function $G(i, j)$ into two $\partial_0 G(i, j) \equiv G(i, 0)G(0, j)$ (The same for $\partial_1 G(i, j) \equiv G(i, 1)G(1, j)$). Such operation inserts an external vertex $0$ ($1$) in the Green's function line. Since the external vertices carry zero momentum for the uniform polarization, the derivatives will not change the internal momenta (the colored loops in Fig.\ref{conservation_law}a). For some vacuum diagrams, the above operations may generate several copies of the same polarization diagrams. However, they have different momentum-loop arrangements and should be kept to maximize the sign cancellation.

The Baym-Kadanoff theory is about the charge conservation of the overall weight of the polarization diagrams. In the imaginary-time and momentum representation, we find that the conservation is already implemented at the diagram topologies level before any explicit integration of the internal variables. This property is crucial for understanding the $U(1)$ symmetry-induced sign cancellation but is far from obvious.  Here we provide a constructive proof of the total charge conservation. A similar approach has been used to prove the Ward identity in quantum electrodynamics in the frequency representation\cite{peskin2018introduction}.

\begin{figure}
    \centering
    \includegraphics[scale=0.265]{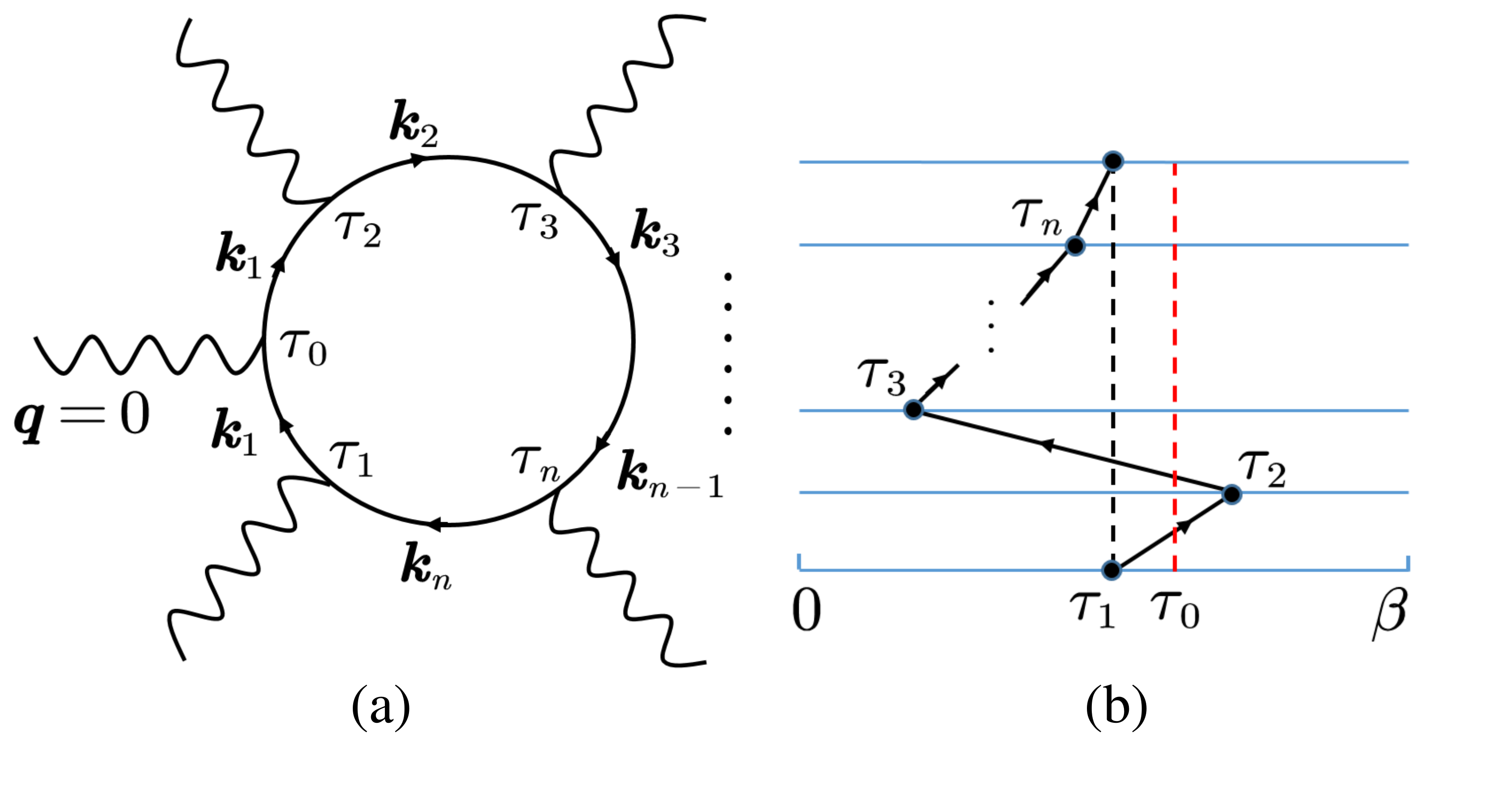}
    \caption{(a) A closed fermion loop with $n$ interaction legs and one external vertex $0$ with an external momentum $\bm{q}=0$. (b) The same fermion loop (before the vertex $0$ is inserted) plotted in a time-ordered manner. The red-dashed line marks the possible time position of the external $\tau_0$. See the main text for more details.}
    \label{loop}
\end{figure}

We consider a fermion loop with $n$ interaction legs as shown in Fig.~\ref{loop}(a). It is the elementary building block of all high-order Feynman diagrams. For example, the polarization diagrams in Fig. \ref{conservation_law} (a) all can be decomposed into two fermion loops. All internal momentum variables ${\bm k}_1, {\bm k}_2, .., {\bm k}_n$ and imaginary-time variables $\tau_1, ..., \tau_{n}$ should not be integrated out. The $\tau$ variables are in the interval $(0, \beta)$ and are unequal to each other. We now pick one Green's function and insert an external vertex $0$ with a given momentum ${\bm q}=0$ and imaginary-time $\tau_0$. Note that $\tau_0$ is the external time, which evolves from $0$ to $\beta$. The operation generates $n$ diagrams. Later, we will show that this is the minimal topological cluster to implement the total charge conservation law. This statement is nontrivial because an individual diagram, as the one in Fig.\ref{loop}(a), violates the total charge conservation.

Without loss of generality, we may choose an arbitrary time order for the internal time variables, as in Fig.\ref{loop}(b). If one inserts the external vertex $0$ between the vertices $i$ and $j$, the two Green's functions $i\rightarrow 0 \rightarrow j$ contribute a weight factor,


\begin{equation}
\begin{aligned}
    G(i,0)G(0,j)&= e^{-\epsilon_{\bm{k}} (\tau_j-\tau_i)} \left\{
    \begin{array}{cc}
        -n_{\bm{k}} (1-n_{\bm{k}})& \tau_0\notin [\tau_i,\tau_j]\\
        (1-n_{\bm{k}})^2 & \tau_j>\tau_0>\tau_i \\ 
        n_{\bm{k}}^2 & \tau_j<\tau_0<\tau_i
    \end{array}  \right.  \\
    &= G(i,j) \left\{
    \begin{array}{cc}
        -n_{\bm{k}} & \tau_j > \tau_i, \tau_0\notin [\tau_i,\tau_j] \\
        1-n_{\bm{k}} & \tau_j < \tau_i,\tau_0\notin [\tau_i,\tau_j] \\
        1-n_{\bm{k}} & \tau_j>\tau_0>\tau_i \\ 
        -n_{\bm{k}} & \tau_j<\tau_0<\tau_i
    \end{array}  \right.  \\
    &\equiv G(i,j) A(\tau_0,\tau_i,\tau_j) \;,
\end{aligned}
\label{eq:vertex_insertion}
\end{equation}
where $A(\tau_0,\tau_i,\tau_j) = 1-n_{\bm{k}}$ or $-n_{\bm{k}}$, depending on the ordering of $\tau_0$,$\tau_i$ and $\tau_j$. It is clear that the diagram weight cannot be a constant of motion as the external time $\tau_0$ evolves from $0$ to $\beta$. 

We next show that the summed weight of all $n$ diagrams is independent of the time evolution of $\tau_0$, thus obeys the charge conservation. We consider the weight 
\begin{equation}
\begin{aligned}
\label{eq:toplogical_conservation}
     \partial_0 G_{loop} & \equiv \partial_0 \left[ G(1,2) G(2,3) \cdots G(n,1) \right]  \\
     & \equiv G(1, 0)G(0, 2)G(2, 3) \cdots G(n, 1) \\
     &+ G(1,2)G(2,0)G(0,3) \cdots G(n, 1) \\
     &+\cdots.
\end{aligned}
\end{equation}
Using Eq.(\ref{eq:vertex_insertion}) and the fact that $A(\tau_0,\tau_i,\tau_j)$ always contains $-n_{\bm{k}}$ despite the time ordering, we derive
\begin{equation}
\begin{aligned}
    &{} \partial_0 G_{loop} =  \\
    &= \left[ A(\tau_0,\tau_1,\tau_2) + A(\tau_0,\tau_2,\tau_3) + \cdots + A(\tau_0,\tau_n,\tau_1) \right] G_{loop} \\
    &=(M-\sum_{i=1}^n n_{\bm{k}_i}) G_{loop}\;,
\end{aligned}
\label{eq:loop_weight}
\end{equation}
where the integer constant $M$ is the number of backward-propagating Green's functions, and the index $i$ sums over all Green's functions. Both quantities should be counted with the fermion loop in the absence of the external vertex $0$. It is therefore independent of $\tau_0$ evolution. 

To prove Eq.(\ref{eq:loop_weight}), one may start with $\tau_0=\beta$ so that $\tau_0$ is larger than all internal imaginary-time variables. One can calculate the factor $(M-\sum_{i=1}^n n_{\bm{k}_i})$ using the first two conditions in Eq.(\ref{eq:vertex_insertion}). We then prove that this factor does not change as $\tau_0$ decreases from $\beta$ to $0$. Without loss of generality, we assume $\tau_2$ is the largest among the remaining time variables. As $\tau_0$ evolves from $\tau_2+0^+$ to $\tau_2-0^+$, the weight of $G(1, 2)$ and $G(2, 3)$ abruptly changes. However, their combined weight remains the same, since two Green's functions must propagate in the opposite temporal direction. As $\tau_0$ further decreases, $\tau_0$ intersects with Green's functions, which are always paired. As a result, the combined weight never changes. We thus conclude that the summed weight of $n$ diagrams must be a constant of motion as in Eq.(\ref{eq:loop_weight}). The derivation can also be generalized to generic charge conservation law with $\bm q\neq 0$. 

\begin{figure}
\centering
    \includegraphics[scale=0.55]{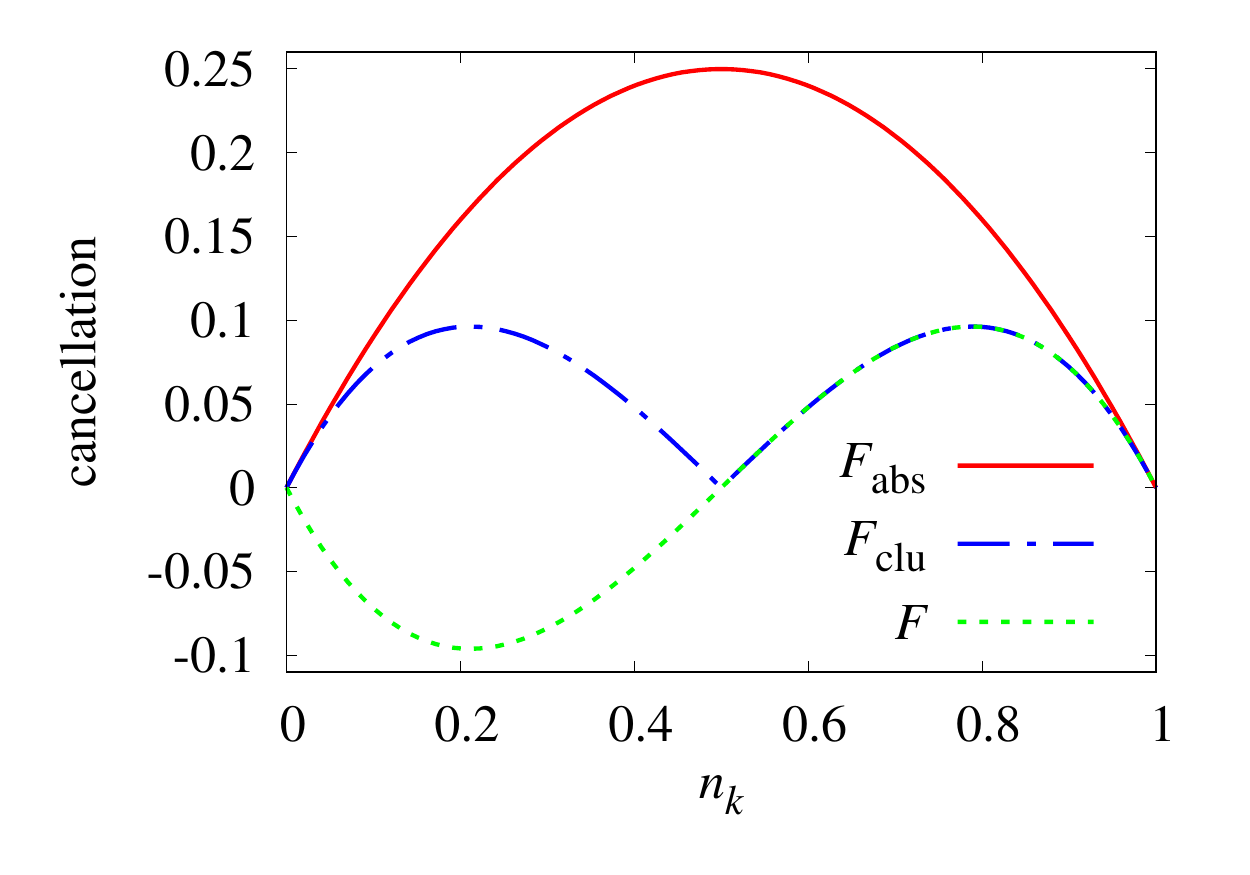}
    \vspace*{-5mm}
    \caption{Sign cancellation between two diagram topologies with a weight $w_1(k)$ and $w_2(k)$, respectively. We plot $w_1(k)+w_2(k)$ (dashed green), $|w_1(k)+w_2(k)|$ (blue) and $|w_1(k)|+|w_2(k)|$ (red) as functions of the internal momentum $k$. They indicate the sign cancellation from different mechanisms (See the main text). 
    }
     \label{cancellation_ex}
\end{figure}

How does the charge conservation lead to the topological sign cancellation? We show it with an inspiring example. Consider the $n=2$ fermion loop and assume $\tau_1 \neq \tau_2$. Most of the diagram weights are associated with small momenta on interaction lines. For clarity, we will simply set all interaction momenta to be zero. This fixes the internal momenta ${\bm k}_1={\bm k}_2={\bm k}$ so that ${\bm k}$ is the only variable in the system. 

Inserting the external vertex $0$ ($\tau_0$ not equal to $\tau_1$ and $\tau_2$) generates two diagrams $0\rightarrow 1 \rightarrow 2 \rightarrow 0$ and $0\rightarrow 2 \rightarrow 1 \rightarrow 0$, which have weights $w_1(k)=(1-n_{\bm k})^2(-n_{\bm k})$, and  $w_2(k)=(1-n_{\bm k})(-n_{\bm k})^2$. The summed weight is thus $w_1(k)+w_2(k)=(-n_{\bm k})(1-n_{\bm k})(1-2n_{\bm k})$. It is independent of the external time as expected. 

We now study the sign cancellation in these two diagrams. With the definition in Sec. \ref{sec:sign_cancellation}, the physical, cluster and absolute weight functions are given by

\begin{align}
    F&=\int \frac{d \bm k}{(2\pi)^2} \left\{ w_1(k)+w_2(k) \right\}, \\
    F_{clu}&=\int \frac{d \bm k}{(2\pi)^2} \left\{|w_1(k)+w_2(k)| \right\}, \label{eq:clu_weight}\\
    F_{abs}&=\int \frac{d \bm k}{(2\pi)^2}\left\{ |w_1(k)|+|w_2(k)| \right\}.
\end{align}

 In Fig. \ref{cancellation_ex}, we plot three integrands as functions of the momentum $k$. Two mechanisms of sign cancellation are observed. The first is the cancellation $|w_1(k)+w_2(k)|<|w_1(k)|+|w_2(k)|$ caused by different diagram topologies. It implies $F_{clu} < F_{abs}$. The second mechanism is implied by the anti-symmetric structure of $w_1(k)+w_2(k)$ near the Fermi momentum. It indicates the physical weight $F \ll F_{clus}$ since $F$ is further suppressed by momentum symmetrization operation $k \rightarrow k_F-k$. Both are responsible for the small constant of motion in Fig.\ref{4_diagram_group}.

Similar to the case of crossing symmetry, the $U(1)$-symmetry-induced sign cancellation is also useful to reduce the sign problem in the conventional DiagMC algorithm\cite{KunChen2019}. In the next Sec.\ref{sec:algorithm}, we provide the protocol to classify Hugenholtz diagrams into conserving groups. In this algorithm, one samples the diagrams with the weight function,
\begin{equation}
    F_{\rm clu}^n=\int[{\rm d}\bm x]^{n} \sum_{\mathscr{C}_n} \left| \sum_{\mathscr{H}_n \in \mathscr{C}_n} h(n,\mathscr{H}_n,\bm x)\right|.
\end{equation}
where $h(n,\mathscr{H}_n,\bm x)$ is the amplitude of Feynman diagrams generated from Hugenholtz topology $\mathscr{H}_n$ which belong to the conserving group $\mathscr{C}_n$. Since each group in this scheme respects the topological conservation law as defined in Eq.\eqref{eq:toplogical_conservation}, we expect $F_{clu}^n/F_{Huge}^n \ll 1$. Indeed, with the polarization of the model \eqref{toymodel}, as shown in Fig.\ref{cancellation}(b), we find this ratio decays approximately exponentially with diagram orders. Compared to the Monte Carlo scheme with Hugenholtz diagrams, the new scheme further improves the efficiency by a factor $\gamma=(F_{clu}^n/F_{Huge}^n)^2$ as shown in the inset of Fig.\ref{hugenholz}b. Combining the Hugenholtz and conserving grouping schemes, the overall efficiency gain compared to the conventional DiagMC is the factor $\gamma=(F_{abs}^n/F_{clu}^n)^2$. We plot this quantity in the inset of Fig.\ref{cancellation}. At the order $6$, we find this factor is of the magnitude $10^5$. Note that in our algorithm, we have not incorporated the momentum symmetrization operation, as discussed in Fig.\ref{cancellation_ex}. We expect that it will further alleviate the sign problem, and we plan to implement it in future publications.

 


\section{Cluster Diagrammatic Monte Carlo Algorithm}
\label{sec:algorithm}
\subsection{Diagram Grouping}
Here we summarize the algorithm to group the diagrams. To ensure massive sign cancellation, one must choose proper topologies as well as internal variables. We will follow the symmetry arguments proposed in the previous section to achieve this goal. We illustrate the key steps as follows:

i) Generate all order-$n$ Hugenholtz vacuum diagrams for the generating functional. A fast algorithm is developed in Ref \cite{tag2017automatic}. For each vacuum diagram, one then labels $2n$ time variables and chooses $n + 1$ independent momentum loops. An example of a second-order diagram can be found in Fig.~\ref{conservation_law} (a), where we label the independent momentum loops with different colors and the imaginary-time variables with numbers. All diagrams must share the same set of variables so that they can be sampled altogether by a Monte Carlo algorithm. To maximize efficiency, it is also essential to choose all loop variables to be fermionic. The main contributions of all integrals come from a thin shell near the Fermi momentum. 

ii) For each order-$n$ vacuum diagram, we generate a group of order-$n+1$ polarization diagrams featuring both the global $U(1)$ symmetry and the crossing symmetry (if possible). This is achieved in two steps: one first inserts two external vertices $0$ and $1$ to the Green's functions in the vacuum diagram in all possible ways, as shown in Fig.~\ref{conservation_law} (a). One then expands each Hugenholtz diagram into $2^n$ Feynman diagrams, meanwhile, eliminates one-interaction-reducible diagrams. In both steps, the derived diagram inherits the internal-variable arrangement of the parent diagram (In the first step, one may need to add one external momentum loop variable if transfer momentum between two external vertices is nonzero). 

We symmetrize the external vertices $0$ and $1$, which reduces the diagram number by two. Nevertheless, the above operations may generate multiple copies with the same topology but the different internal-variable arrangement, as shown in Fig.~\ref{conservation_law} (a). Those copies are required by the symmetry and the conservation law. 

In the end, we point out that the above scheme achieves better sign cancellation than the scheme proposed in Ref.\cite{KunChen2019}. One of the most significant differences is that the previous scheme only keeps one copy for each topology, thus violating the symmetry and resulting in considerably less topological sign cancellation.

\subsection{Monte Carlo Algorithm}

We will only briefly explain the protocol because we use the standard Monte Carlo integration algorithm to calculate the diagram weights.

The configuration space in our algorithm consists of the diagram order $n$, imaginary-time variables and momentum loop variables. The probability density of the order $n\ge 1$ contribution is given by the weight function,
\begin{equation}
    \rho(n\ge 1, \bm x) \propto \sum_{\mathscr{C}_n} \left| \sum_{\mathscr{H}_n \in \mathscr{C}_n} h(n,\mathscr{H}_n,\bm x)\right|, 
\end{equation}
where $\bm x$ represents all internal variables, and $h(n,\mathscr{H}_n,\bm x)$ is the amplitude of an $n$-order Feynman diagram belonging to the Hugenholz topology $\mathscr{H}_n$, and $\mathscr{C}_n$ is the collection of order $n$ conserving groups.

To normalize the integral, we also include a special zeroth order ``diagram",
\begin{equation}
    \rho(n=0) \propto 1,
\end{equation}
which is simply a normalization constant without any internal variables.

The minimal set of Monte Carlo updates includes three operations: increase or decrease diagram order by one (add or remove a pair of internal variables), and randomly select an internal variable to update. The standard Metropolis algorithm is used for the acceptance probability. By adjusting the re-weighting factor of each order, we find all updates are quite efficient. 

A detailed discussion of the statistical error-bar estimation is given in Appendix \ref{errorbar}.




\section{conclusion}
In conclusion, we reveal the fermionic sign structure of high-order Feynman diagrams constrained by the crossing symmetry and the charge conservation law, which are two universal features of many-fermion systems. As a by-product, we prove that the charge conservation law is a topological property of Feynman diagrams in the imaginary-time/momentum representation. With modern numerical techniques, we investigate the significance of sign cancellations from the crossing symmetry and the conservation law.

The diagrammatic sign structure knowledge is used to optimize the internal-variable arrangements and construct the sign-canceled diagram groups. We then propose a cluster Diagrammatic Monte Carlo algorithm, which samples the sign-canceled diagram groups instead of individual diagrams. Compared to the conventional algorithm, we achieve an efficiency boost up to a factor $\sim 10^5$ at the diagram order $6$.

Besides the numerical application, identifying the sign-canceled groups provides important hints in constructing the relevant perturbative effective field theory of many-fermion systems. The diagrams in Fig.\ref{4_diagram_group}, which forms a sign-canceled group, provides a particularly relevant example. The massive sign cancellation between those diagrams results in an emergent ``unreasonably'' small number, invalidating the conventional power-counting argument. It indicates that the relevance of perturbation in fermionic systems is better discussed in the context of sign-canceled groups rather than individual diagrams.

In the end, we summarize open questions for future work. There are questions associated with the charge conservation law in the topological structure of diagrams. Are there similar properties for other conservation laws (e.g., energy, momentum, and angular momentum)? Does the type of internal variables (space versus momentum, time versus frequency) matter? One probably needs to formulate a stronger version of the celebrated Baym-Kadanoff theory to address the above questions. There are also technical problems. For example, in Sec.\ref{sec:conservation}, we point out that the momentum symmetrization operation $k\rightarrow k_F-k$ leads to a significant sign cancellation as indicated by Fig.~\ref{cancellation_ex}. However, it is not clear how to implement it in the Monte Carlo algorithm to alleviate the sign problem.

\begin{acknowledgments}
	The authors would like to thank Nikolay Prokof’ev, Boris Svistunov, and Zhiyuan Yao for stimulating discussion and important input. B.-Z. Wang, P.-C. Hou and Y. Deng were supported by the National Natural Science Foundation of China (under Grant No. 11625522) and by the National Key R\&D Program of China (under Grants No. 2016YFA0301604 and No. 2018YFA0306501). K. Chen was supported by the Simons Collaboration on the Many Electron Problem and K. Haule was supported by NSF DMR-1709229.
\end{acknowledgments}


\bibliographystyle{apsrev}
\bibliography{references}

\appendix
\setcounter{equation}{0}
\label{errorbar}
\section{Statistic Error Estimation in Diagrammatic Monte Carlo Algorithms}
In our diagrammatic Monte Carlo algorithm, the configuration spaces consist of two sub-spaces                                                                                                                    : the physical diagrams with orders $n\ge 1$ and the normalization ``diagram'' with the order $n=0$. The weight function of the latter, $g({\bm x})$, should be simple enough so that the integral $G=\int g({\bm x}) d\bm x$ is explicitly known. In our algorithm we use a constant $g(\bm x) \propto 1$ for simplicity. In this setup, the physical diagram weight $F$ can be calculated with the equation
\begin{equation}
\label{FG}
    F=\frac{F_{\rm MC}}{G_{\rm MC}} G,
\end{equation}
where the MC estimators $F_{\rm MC}$ and $G_{\rm MC}$ are measured with 
\begin{equation}
\begin{aligned}
\label{sigma}
  F_{\rm MC} &=\frac{1}{N} \left[ \sum_{i=1}^{N_f} \frac{f(\bm{x}_i)}{\rho_f(\bm{x}_i)} + \sum_{i=1}^{N_g} 0 \right] \\
  G_{\rm MC} &=\frac{1}{N} \left[\sum_{i=1}^{N_f} 0 + \sum_{i=1}^{N_g} \frac{g(\bm{x}_i)}{\rho_g(\bm{x}_i)}  \right]\;.
\end{aligned}
\end{equation}
For simplicity, here we have suppressed the labels such as the order $n$ and the topological index $\xi_n$ in the diagram weight $f(\bm{x}_i)$, so that the total amplitude is a simple integral $F = \int f({\bm x}) d{\bm x}$. The probability density of a given configuration is proportional to $\rho_{f}(\bm x)=|f(\bm x)|$ and $\rho_{g}(\bm x)=|g(\bm x)|$, respectively. In the total MC steps $N$, the physical diagrams are sampled for $N_f$ times, and the normalization ``diagram'' are for $N_g$ times.

Now we estimate the statistic error. According to the propagation of uncertainty, the variance of $F$ in Eq.\eqref{FG} is given by
\begin{equation}
    \sigma^2_F = \left( \frac{G}{G_{\rm MC}} \right)^2 \sigma_{F_{\rm MC}}^2 +  \left( \frac{G F_{\rm MC}}{G^2_{\rm MC}} \right)^2 \sigma_{G_{\rm MC}}^2,
    \label{propagation}
\end{equation}
where $\sigma_{F_{\rm MC}}$ and $\sigma_{G_{\rm MC}}$ are variance of the MC integration $F_{\rm MC}$ and $G_{\rm MC}$, respectively. In the Markov chain MC, according to definition the variance of $F_{\rm MC}$ can be written as 
\begin{align}
\sigma^2_{F_{\rm MC}} &= \frac{1}{N} \left[ \sum_{i}^{N_f} \left( \frac{f(\bm{x}_i)}{\rho_f(\bm{x}_i)}- \frac{F}{Z}\right)^2 +\sum_{j}^{N_g} \left(0-\frac{F}{Z} \right)^2  \right] \nonumber \\
&= \int \left( \frac{f(\bm x)}{\rho_f(\bm x)} - \frac{F}{Z} \right)^2 \frac{\rho_f(\bm x)}{Z} {\rm d}\bm x + \int \left( \frac{F}{Z} \right)^2 \frac{\rho_g(\bm x)}{Z}{\rm d}\bm x \notag  \nonumber \\ 
&=  \int \frac{f^2(\bm x)}{\rho_f(\bm x)} \frac{{\rm d}{\bm x}}{Z} -\frac{F^2}{Z^2} \; .
\end{align}
Here $Z=Z_f+Z_g$ and $Z_{f/g}=\int \rho_{f/g}({\bm x})d{\bm x}$ are the partition sums of the corresponding configuration spaces. Due to the detailed balance, one has $Z_f/Z_g=N_f/N_g$. 

Similarly, the variance of $G_{\rm MC}$ can be written as 
\begin{align}
 \sigma^2_{G_{\rm MC}}=  \int \frac{g^2(\bm x)}{\rho_g(\bm x)} \frac{{\rm d}{\bm x}}{Z} - \frac{G^2}{Z^2} \; .
\end{align}
By substituting $\rho_{f}(\bm x)=|f(\bm x)|$ and  $\rho_{g}(\bm x)=|g(\bm x)|$, the variances of $F_{\rm MC}$ and $G_{\rm MC}$ are given by
\begin{equation}
\begin{aligned}
    \sigma^2_{F_{\rm MC}}&= \frac{1}{Z^2} \left( Z Z_f - F^2 \right)\;, \\
    \sigma^2_{G_{\rm MC}}&= \frac{1}{Z^2} \left( Z Z_g - G^2 \right)\;.
\end{aligned}
\end{equation}
With the Eq.~\ref{propagation}, we derive the variance of $F$ as
\begin{equation}
\begin{aligned}
\sigma^2_F &= Z\left(Z_f+\frac{F^2}{G^2} Z_g\right) - 2F^2 \\
		&\approx Z_f^2\;.
\end{aligned}
\end{equation}
The approximations are based on the facts that $F\ll Z$ (massive sign cancellation), and $Z_f \gg Z_g$. The final expression is simple but heuristic: the variance of $F$ only depends on the partition sum $Z_f$. The statistical error after $N$ MC steps is
\begin{equation}
    \epsilon \sim \frac{\sigma}{N^{1/2}} \frac{Z_f}{N^{1/2}}.
\end{equation}



\end{document}